\documentclass{svjour2}                    % onecolumn
\usepackage{amsmath,amssymb}
\smartqed  % flush right qed marks, e.g. at end of proof
\usepackage{graphicx}
\usepackage{color} 

\begin{document}

\title{Review: Observation of Majorana Bound States at a Free Surface of $^3$He-B}

%\subtitle{Do you have a subtitle?\\ If so, write it here}

%\titlerunning{Short form of title}        % if too long for running head

\author{Hiroki Ikegami         \and
          Kimitoshi Kono 
}

%\authorrunning{Short form of author list} % if too long for running head

\institute{H. Ikegami  \and K. Kono \at
             RIKEN Center for Emergent Matter Science (CEMS), Wako, Saitama 351-0198, Japan\\
              \email{hikegami@riken.jp}           %  \\
%             \emph{Present address:} of F. Author  %  if needed
 \and
K. Kono \at present address:
Department of Electrophysics, NCTU, 1001 Ta-Hsueh Rd, Hsinchu, Taiwan ROC\\
}

\date{Received: date / Accepted: date}
% The correct dates will be entered by the editor

\maketitle

\begin{abstract}
The $p$-wave superfluid $^3$He is a textbook example of topological superfluids.
Among its multiple superfluid phases, the B phase ($^3$He-B) is known as a topological state protected by  time-reversal symmetry.
One of the important topological features of $^3$He-B is the formation of bound states at its surface.
Notably, such surface Andreev bound states are predicted to be Majorana fermions, i.e., their antiparticles are identical to their particles. 
Because of the well-elucidated bulk properties of the superfluid $^3$He owing to its cleanliness, $^3$He-B provides an ideal platform to pursue Majorana fermions in condensed matter systems.
In this article, we review recent investigations of surface Andreev bound states by the mobility of ions trapped below a free surface of $^3$He-B.
The free surface is an ideal surface providing a specular boundary condition; 
The surface Andreev bound states formed there are expected to be Majorana fermions with a well-defined energy spectrum.
We show that the temperature and depth dependences of the experimentally obtained mobility of negative ions [H. Ikegami {\it et al.,} {\it J. Phys. Soc. Jpn.} {\bf 82} 124607 (2013)] are {\it quantitatively} reproduced by a theoretical study that includes scattering of the surface Andreev bound states [Y. Tsutsumi, {\it Phys. Rev. Lett.} {\bf 118} 145301 (2017)].
This quantitative agreement unambiguously demonstrates the experimental detection of surface Andreev bound states.
We also discuss the future prospects of the Majorana physics in $^3$He-B.

% \PACS{PACS code1 \and PACS code2 \and more}
% \subclass{MSC code1 \and MSC code2 \and more}
\end{abstract}

\section{Introduction}

One of the long-pursued but yet-to-be discovered particles are Majorana fermions.
They are hypothetical elementary particles \cite{Elliott2015}, that satisfy the condition that their antiparticles are identical to their particles.
Although such particles have not been found in particle physics, recent theoretical studies surprisingly suggest that Majorana fermions can be generated as emergent excitations in condensed matter systems \cite{Read2000,Qi2011,Sato2016,Sato2017,Alicea2012,Beenakker2013,Aguado2017}.
Thus, the naturally arising questions are whether or not such exotic particles really exist in nature and what novel properties they have if they exist.

Fermi superfluids and superconductors are natural candidates for systems that host Majorana fermions.
In these systems, quasiparticles (QPs) are described as superpositions of a particle and a hole state, and therefore the QPs could be identical to their antiparticles if certain conditions are satisfied \cite{Read2000,Qi2011,Sato2016,Sato2017,Alicea2012,Beenakker2013,Aguado2017}.
In fact, Majorana fermions are predicted to appear as emergent excitations at a surface or in a vortex of topological superfluids/superconductors.  
Majorana fermions have attracted considerable attention for the last decade not only because of fundamental interest in their properties but also for their potential application to fault-tolerant quantum computing, and much experimental effort has been devoted to the search for Majorana fermions  \cite{Sato2016,Sato2017,Beenakker2013}.

The principal systems that host Majorana excitations are odd-parity topological superfluids/superconductors.
The $p$-wave superfluid $^3$He is a prime candidate.
For superconductors, there are several candidates which are now conceived to have a nontrivial topology in the bulk order parameters \cite{Yonezawa2016}, such as Sr$_2$RuO$_4$ \cite{Maeno2012}, the B phase of UPt$_3$ \cite{Izawa2014,Goswami2015}, Cu$_x$Bi$_2$Se$_3$ \cite{Sasaki2015}, and Sn$_{1-x}$In$_x$Te \cite{Sasaki2015}.
In some of these superconductors, the presence of edge or surface Andreev bound states has been reported by tunneling spectroscopy measurements (Sr$_2$RuO$_4$ \cite{Kashiwaya2011}, Cu$_x$Bi$_2$Se$_3$ \cite{Sasaki2011}, and Sn$_{1-x}$In$_x$Te \cite{Sasaki2012}), suggesting topologically nontrivial states in the bulk.
However, the bulk order parameters have not yet been firmly established for these superconductors.
Other intensively studied systems is a nanowire-superconductor hybrid system, where a topological superconducting state is induced in the nanowire \cite{Oreg2010,Lutchyn2010}.
Zero-energy states have been observed at the ends of nanowires by tunneling spectroscopy \cite{Mourik2012,Das2012,Deng2012,Deng2016,Albrecht2016}, but there is an unsettled question as to whether they are trivial or nontrivial zero modes at present \cite{Lee2014}.

In order to make clear understanding of properties of Majorana fermions, experiments in well-elucidated topological systems are essential.
The superfluid $^3$He \cite{Vollhardt90,Dobbs00,Volovik2003} provides an ideal platform for exploring the nature of Majorana fermions because its bulk properties are well elucidated and clear experimental results are obtainable owing to the cleanliness of liquid $^3$He.
One of the superfluid phases, called the B phase ($^3$He-B), is recognized as a topological superfluid protected by time-reversal symmetry \cite{Mizushima2016,Schnyder2008,Chung2009}.
On the surface of $^3$He-B, Majorana fermions are expected to emerge as surface Andreev bound states \cite{Schnyder2008,Chung2009,Qi2009,Volovik2009_1}.

The surface Andreev bound states in $^3$He-B have been observed in the last decade by transverse acoustic impedance and other methods \cite{Aoki2005,Okuda2012,Murakawa2009,Murakawa2011,Choi2006}.
Their observations are very important because they verify that the nontrivial topology of $^3$He-B generates such bound states.
In these studies, however, the employed surfaces have more or less microscopic roughness,  which prevents understanding of detailed properties of surface Andreev bound states.
At a smooth surface called a specular surface, on the other hand, surface Andreev bound states with a well-defined energy spectrum are formed, and exotic features of Majorana fermions should be unambiguously elucidated. 
A specular surface is realized at a free surface (see Sect. \ref{sect:specularity}).
In order to detect Majorana fermions formed at a free surface, Ikegami {\it et al.} measured the mobility of ions trapped below a free surface of $^3$He-B in 2013 \cite{Ikegami2013-2}. 
Recently, the experimental mobility was quantitatively reproduced by a calculation by Tsutsumi that included contributions from surface Andreev bound states \cite{Tsutsumi2017}.
This quantitative agreement unambiguously demonstrates the experimental detection of surface Andreev bound states at a free surface.
In this review, we describe such recent progress with the aim of providing a clear view of the detection of surface Andreev bound states from the mobility of ions.

%-----------------------------------------------------------
\section{Majorana Fermions Formed at a Surface of $^3$He-B}
\label{Majorana_3He}

The superfluid $^3$He is a $p$-wave pairing state with orbital angular momentum 1 and spin angular momentum 1.
In $^3$He-B, the condensate is composed of an admixture of Cooper pairs with three substates of the orbital angular momentum ($L_z=$ $-$1, 0, 1) and three substates of the spin ($S_z=$ $-$1, 0, 1), and 
the orbital angular momentum $\mbox{\boldmath $L$}$ and the spin angular momentum $\mbox{\boldmath $S$}$ couple with each other such that the twisted total angular momentum becomes zero [$\mbox{\boldmath $L$}+R(\mbox{\boldmath $\hat n$},\theta)^{-1}\mbox{\boldmath $S$}=0$, where $R(\mbox{\boldmath $\hat n$},\theta)$ is the rotation matrix].
The order parameter $\mbox{\boldmath $d$}$ is then described as \cite{Vollhardt90,Dobbs00}
\begin{equation}
\mbox{\boldmath $d$} (\mbox{\boldmath $\hat k$}) = \Delta_B R(\mbox{\boldmath $\hat n$},\theta) \mbox{\boldmath $\hat k$} ,
\label{eq:OP} 
\end{equation}
where $\Delta_B$ is the energy gap of $^3$He-B and $\mbox{\boldmath $\hat k$}$ is a unit vector in momentum space. 
This state breaks the relative spin-orbit rotation symmetry because of the coupling of $\mbox{\boldmath $L$}$ and $\mbox{\boldmath $S$}$.
This state is however invariant under time inversion operation $\hat T$ because $\hat T \mbox{\boldmath $d$} (\mbox{\boldmath $\hat k$}) = -\mbox{\boldmath $d$}^* (-\mbox{\boldmath $\hat k$})= \mbox{\boldmath $d$} (\mbox{\boldmath $\hat k$})$.

One of the notable phenomena expected in $^3$He-B is the formation of surface Andreev bound states in the region within the coherence length from the surface.
In 1981, Buchholtz and Zwicknagl first pointed out that $s$-wave and $p$-wave superconductors respond to a surface differently and showed that Andreev bound states are formed at the surface of $^3$He-B \cite{Buchholtz1981}. 
Since then, a number of theoretical investigations not only for a specular surface but also for surfaces in the presence of microscopic roughnesses have been carried out for $p$-wave \cite{Nagato1998,Zhang1987,Zhang1988,Nagai2008} and $d$-wave pairing states \cite{Hu1994,Matsumoto1995,Kashiwaya1995}.
Basically, surface Andreev bound states are formed as a consequence of the interference of an incident wavefunction and a reflected wavefunction of a QP \cite{Ohashi1996}.

The surface Andreev bound states of  $^3$He-B are intimately related to the non-trivial topology of the bulk $^3$He-B order parameter.
In the superfluid $^3$He, investigations of the topological aspects  have already begun as early as the 1980's \cite{Salomaa1988}.
After that, a number of unusual behaviors have been uncovered by using the concept of topology \cite{Volovik2003}, especially in the last decade, along with the drastic progress in the topological physics in condensed matter systems \cite{Qi2011,Sato2016,Sato2017,Beenakker2013,Mizushima2016,Hasan2010,Ando2013}.
Now, $^3$He-B is recognized as a topological state protected by time-reversal symmetry \cite{Mizushima2016,Schnyder2008,Chung2009,Mizushima2015}.
In general, the nontrivial topology of a bulk gives rise to formation of gapless bound states on its surface owing to the so-called bulk-surface correspondence.
In $^3$He-B, gapless surface Andreev bound states are formed as a result of the broken symmetry of the bulk $^3$He-B order parameter.
Because of equal-weight superpositions of a particle and a hole state, surface Andreev bound states satisfy the Majorana condition as shown below \cite{Schnyder2008,Chung2009}.

To see basic properties of surface Andreev bound states, we consider solutions of the Bogoliubov-de Gennes equation near a surface.
We take the $z$-axis normal to the surface of $^3$He-B which fills the region at $z>0$.
We approximate the perpendicular and the parallel components of the order parameter as $\Delta_\bot(z)=\Delta_B \tanh (\kappa z)$ and $\Delta_\parallel (z)=\Delta_B$ near the surface respectively ($\kappa  = \Delta _B /\hbar v_F$) \cite{Tsutsumi2012-2}, which provides a good approximation for the suppression of the order parameter near the surface (see Appendix in Ref. \cite{Tsutsumi2012-2}).
The field operator for the surface Andreev bound states is then given by \cite{Tsutsumi2017}
\begin{equation}
\begin{array}{rr}
 \left[ {\begin{array}{*{20}c}
   {\psi _ \to  ({\bf{r}})}  \\
   {\psi _ \leftarrow  ({\bf{r}})}  \\
   {\psi _ \to ^\dag  ({\bf{r}})}  \\
   {\psi _ \leftarrow ^\dag  ({\bf{r}})}  \\
\end{array}} \right] & = \sum \limits_{\bf{k}} {A_k \frac{1}{\cosh (\kappa z)} \sin (k_ \bot  z)\left( {\hat \gamma _{{\bf{k}} } e^{i{\bf{k}}_\parallel \cdot {\bf{r}} }  + \hat \gamma _{{\bf{k}} }^\dag  e^{ - i{\bf{k}}_\parallel  \cdot {\bf{r}} } } \right)}  \\ 
  & \times \left[ {\begin{array}{*{20}c}
   {\cos \left[ {(\phi _k  + \pi /2)/2} \right]}  \\
   {\sin \left[ {(\phi _k  + \pi /2)/2} \right]}  \\
   {\cos \left[ {(\phi _k  + \pi /2)/2} \right]}  \\
   {\sin \left[ {(\phi _k  + \pi /2)/2} \right]}  \\
\end{array}} \right] . \\
 \end{array}
\label{eq_SABS}
\end{equation}
Here $\hat \gamma _{{\bf{k}} }^\dag$ ($\hat \gamma _{\bf{k}}$) is a Fermion creation (annihilation) operator, ${\bf{k}}=({\bf{k}}_\parallel,k_\bot)$, ${\bf{k}}_\parallel$ is the momentum parallel to the surface,  $k_\bot^2=k_F^2-{\bf{k}}_\parallel ^2$,  $A_k$ is a normalization constant, and $\to $ ($\leftarrow$) represents the spin $\sigma$ with a quantization axis taken in the $x$-direction ($\hbar$ is Planck's constant divided by 2$\pi$, $v_F$ is the Fermi velocity, and $k_F$ is the Fermi wavenumber).
These states localize in the vicinity of the surface with the length scale of $\kappa^{-1}=3.6\xi_0$, where $\xi_0=\hbar v_F/2\pi k_BT_c \sim$ 80 nm is the coherence length ($k_B$ is Boltzmann's constant and $T_c$ is the superfluid transition temperature).
The modes described in Eq. (\ref{eq_SABS}) have a gapless linear dispersion relation with energy $E$ given by $E= \hbar c \left| \bf{k}_\parallel \right|$ with $c=\Delta_B/(\hbar k_F)$.
This dispersion relation is called Majorana cone. 
As shown in Refs. \cite{Chung2009,Mizushima2015,Mizushima2016}, the surface Andreev bound states are described by the massless two-dimensional (2D) Dirac Hamiltonian.
From the linear dispersion relation, the local density of states of the surface Andreev states is obtained as $N_s(E,z)= \frac{\pi}{4} N_0 \frac{E}{\Delta_B} \frac{1}{[\cosh (\kappa z)]^{2}}$, where $N_0$ is the density of states at Fermi energy in normal $^3$He \cite{Tsutsumi2012-2}.

The remarkable feature of surface Andreev bound states is that they are described as Majorana fermions.
In fact, they behave as relativistic QPs described by the massless 2D Dirac Hamiltonian \cite{Chung2009,Mizushima2015,Mizushima2016} and satisfy the Majorana condition, $\psi _ \sigma ({\bf{r}}) = \psi _  \sigma ^\dag  ({\bf{r}})$, as seen in Eq. (\ref{eq_SABS}).

\begin{figure}
\begin{center}
\includegraphics[width=0.7\linewidth,keepaspectratio]{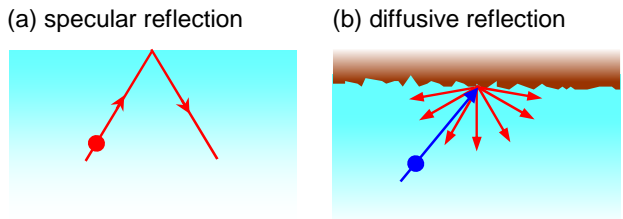}
\end{center}
\caption{\label{surface_condition} (Color online) Schematic pictures of (a) specular reflection and (b) diffusive reflection for a QP in normal state.
}
\end{figure}

The properties of surface Andreev bound states are strongly affected by nature of the surface.
In the case of an ideal surface called specular surface [Fig.  \ref{surface_condition}(a)], a QP in normal state is reflected in the angle same to the incident angle.
In fact, the wavefunction described by Eq. (\ref{eq_SABS}) is the solution for the specular surface;
 the modes in Eq. (\ref{eq_SABS}) are eigenfunction of the momentum parallel to the surface ${\bf{k}}_{\parallel}$.
(Note that the momentum parallel the surface is conserved in the specular reflection.) 
Such a specular surface is realized at a free surface of liquid $^3$He, as shown in Sect. \ref{sect:specularity}.
Another important surface often used in experimental studies is a wall of a solid with which liquid $^3$He is in contact [Fig.  \ref{surface_condition}(b)].
In this case, irregularities in atomic scale ($\sim k_F^{-1}$) existing on the surface make QPs reflected in random directions.
In this diffusive reflection, ${\bf{k}}_{\parallel}$ is not conserved, which makes the spectra of surface Andreev bound states to be broadened and shifted significantly \cite{Nagato1998,Nagai2008}.
Therefore, surface Andreev bound states in this case no more have a well-defined Majorana cone. 
Note that, in the diffusive reflection, the momentum  on the order of $p_F (=\hbar k_F)$ parallel to the surface is exchanged between the wall and the QP.

Experimentally, surface Andreev bound states formed on solid surfaces have been observed so far.  
After the first detection by transverse acoustic impedance measurement in 2005 \cite{Aoki2005}, they have been observed not only by this method \cite{Okuda2012,Murakawa2009,Murakawa2011} but also by specific heat \cite{Choi2006}, by the attenuation of transverse sound \cite{Davis2008}, and potentially by damping and critical velocity of a micromechanical oscillator \cite{Zheng2016,Zheng2017}.
In particular, a series of studies using transverse acoustic impedance \cite{Okuda2012,Murakawa2009,Murakawa2011} showed, although indirectly, that the dispersion relation of surface Andreev bound states approaches to the Majorana cone by changing the surface quality from diffusive to partially specular by coating the surface with $^4$He layers; 
The experiments themselves cold not detect the dispersion relation, but the observed acoustic impedance was consistently explained by the theory in which the dispersion relation is assumed to approach to the Majorana cone as increasing specularity.
However, even if the specularity is enhanced by the $^4$He coating, the surface cannot be perfectly specular for a solid surface, which prevents us from quantitative understanding of nature of surface Andreev bound states.
On the other hand, at a specular surface, where surface Andreev bound states should be Majorana fermions having a well-defined linear dispersion, one can reach unambiguous understanding of their properties.

%-----------------------------------------------------------
\section{Specular Nature of Free Surface of Liquid $^3$He}
\label{sect:specularity}

As mentioned in Sect. \ref{Majorana_3He}, Majorana surface Andreev bound states have a linear dispersion relation only at a specular surface.
A specular surface can be realized at a free surface of liquid $^3$He.
The free surface is an interface in contact with $^3$He vapor (essentially a vacuum in the sub-mK region), and there is no impurity or disturbance at the surface, except for thermally excited surface waves.
The thermal surface waves, however, do not degrade the specular nature of reflection.
To make the reflection diffusive, irregularity of the surface at a length scale on the order of the Fermi wavelength $\lambda _F =2\pi/ k_F$ ($=$ 0.82 nm)  is necessary, but the wavelength of thermally excited surface waves is, at 1 mK for example, $\lambda _{\rm th} =2\pi (\alpha_3/\rho_3)^{1/3}(\hbar /k_BT)^{2/3} \sim$ 0.3 $\mu$m, much longer than $\lambda _F$.
(Here we used the dispersion relation of capillary waves $\omega=(\alpha_3/\rho_3)^{1/2}k^{3/2}$ with frequency $\omega$ and wavenumber $k$, where $\alpha_3$ and $\rho_3$ are the surface tension and density of liquid $^3$He, respectively.)
Therefore, the free surface can be regarded as smooth at the atomic scale even in the presence of thermally excited surface waves, suggesting that QPs are reflected specularly.

The specular nature of a free surface of liquid $^3$He has actually been demonstrated via the mobility $\mu$ of a Wigner crystal of electrons floating above a free surface \cite{Book2DE,BookMonarkha}.
In the Wigner crystal phase, which is formed at sufficiently low temperatures (the transition temperature is 0.32 K for an electron density of 1$\times$10$^{12}$ m$^{-2}$ for example), the electrons exert a spatially modulated pressure on the surface, generating a deformation of the surface commensurate with the Wigner crystal called a dimple lattice.
Although the depth of the dimple  $\delta$ is only $\sim$ 0.1 \AA , the dimple lattice significantly affects the mobility of elections parallel to the surface.
In particular, the motion of the Wigner crystal involves the motion of the dimple lattice, which is impeded by a drag force acting on the dimple lattice.
In the ballistic regime realized at low temperatures ($T \lesssim$ 20 mK), the drag force is caused by the reflection of ballistic QPs at the surface. 
In this ballistic regime, experimental studies have revealed that $\mu$ in normal $^3$He is independent of $T$ \cite{Shirahama1997,Monarkha1998}, and $\mu$ in $^3$He-B and $^3$He-A rapidly increases with decreasing $T$ \cite{Shirahama1997,Monarkha1998,Ikegami2006}.
These experimental mobilities are quantitatively explained by the theoretical model of specular reflection \cite{Monarkha1997}, demonstrating the specular reflection of QPs at the free surface.

The quantitative agreement of the mobility with the specular reflection model indicates that the fraction of diffusively reflected QPs should be very small (less than one in 10$^{10}$ of the incident QPs).
This can be understood by noting that even such a small fraction of diffusively-reflected QPs can greatly reduce the mobility. 
To see this, we consider a drag force $dF_D$ acting on a surface element $dS$ of a dimple lattice moving horizontally with velocity $V_0$.
The drag force in the case of specular reflection is given as
\begin{equation}
 dF_D^{(s)}  \sim n_3 p_F V_0 (\delta /a)^2 dS ,
\label{eq:specular} 
\end{equation}
while that for diffusive reflection is 
\begin{equation}
dF_D^{(d)} \sim n_3 p_F V_0 dS ,
\label{eq:diffuive} 
\end{equation}
where $a$ is the period of the dimple lattice ($\sim$ 1 $\mu$m) and $n_3$ is the number density of liquid $^3$He \cite{Monarkha2006} (see also the argument described in Ref. \cite{Ikegami2017}).
Equations (\ref{eq:specular}) and (\ref{eq:diffuive}) suggest that the diffusive reflection generates a drag force that is larger by a factor of $(a/ \delta)^2$ ($\sim$ 10$^{10}$).
Therefore, the mobility should be significantly suppressed even if a tiny fraction of QPs on the order of 10$^{-10}$ undergo conventional diffusive reflection.

%-----------------------------------------------------------
\section{Ions and Experimental Details}
\subsection{Ions in Liquid $^3$He}
\label{sect:ion}

Ions in liquid $^3$He are charged objects at the mesoscopic scale, which have been used to investigate the microscopic properties of liquid $^3$He.
There are two species of ions: a negatively charged ion called an electron bubble and a positively charged ion called a snowball \cite{Fetter1976}.
In this article, we only consider the negative ion.
The negative ion is the state of an electron self-trapped in a small spherical void with a radius of about 1 nm \cite{Fetter1976}.
The negative ion can be generated by injecting an electron into liquid helium, and the void is created as a result of the Pauli exclusion principle between the electron and electrons in the shells of $^3$He atoms.
The size of the void is determined by the balance of the zero-point energy of the confined electron and the surface tension of the void at zero pressure.
The void can be accurately regarded as a hard-sphere potential for QPs \cite{Shevtsov2017} because QPs cannot penetrate inside the void due to a huge repulsive barrier.
The hard-core radius $R$ is estimated to be $R=11.17k_F^{-1}$ ($k_F^{-1}=$ 0.13 nm) \cite{Shevtsov2016} from the low-temperature mobility in normal $^3$He at zero pressure obtained by Ikegami {\it et al.} \cite{Ikegami2013-2,Ikegami2015}.

Transport measurements of the negative ions have been used to uncover unusual aspects of the superfluid $^3$He.
In bulk $^3$He-B, the mobility has revealed that the $p$-wave coherence of QPs causes an unusual reduction in the transport cross section of the ions \cite{Baym1977,Baym1979,Ahonen1978}.
In $^3$He-A, transport of the ions was found to exhibit an anomalous Hall effect \cite{Ikegami2015,Ikegami2013-1} as a result of the skew scattering of QPs by the ions \cite{Shevtsov2016}, which directly demonstrated the breaking of time-reversal symmetry.
(We note that Refs.\cite{Salmelin1989,Salmelin1990} pointed out that the chiral order parameter of $^3$He-A could allow for transverse Hall term in mobility tensor.
However, the formulation described there gave zero Hall term \cite{Shevtsov2016}.)
In this article, we show that the transport properties of the negative ion are also useful for detecting surface Andreev bound states.

\subsection{Ions Trapped below Free Surface}
\label{sect:ion_below_surface}

\begin{figure}
\begin{center}
\includegraphics[width=0.6\linewidth,keepaspectratio]{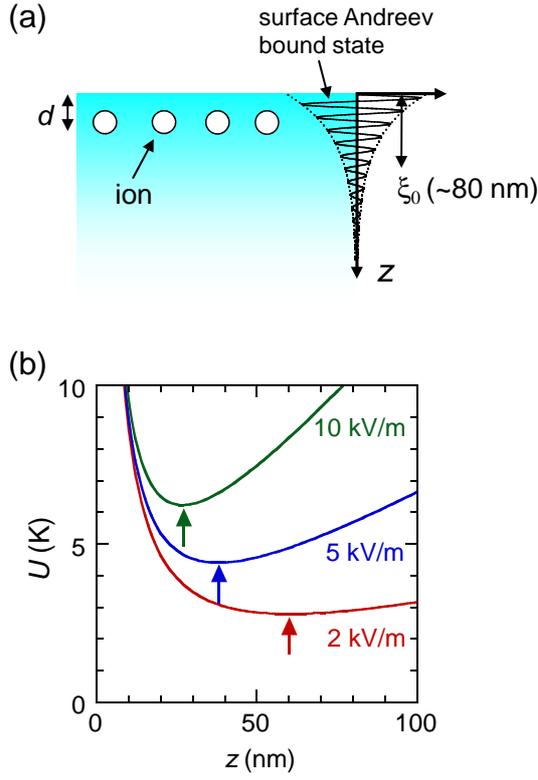}
\end{center}
\caption{\label{trapped_ion} (Color online) (a) Surface Andreev bound states and ions trapped below the free surface. The surface Andreev bound states are formed in the region within $\sim \xi_0 $ ($\sim$ 80 nm) from the surface, and the ions are trapped at a depth that is tunable in the range of 20--60 nm. (b) Trap potential for an ion as a function of distance $z$ from the surface. The curves represent the potential [Eq. (\ref{potential})] at $E_{ \bot }$=2, 5, and 10 kV/m. The arrows indicate the minimum in the potential, at which the ions are trapped. 
}
\end{figure}

Ions can be trapped below the free surface in a potential well $U(z) $ produced by the combination of the image charge of an ion and an external vertical electric field $E_{ \bot }$ \cite{Poitrenaud1972} [Fig. \ref{trapped_ion}(b)]:
\begin{equation}
U(z)  = \frac{{1}}{{16\pi \varepsilon \varepsilon _0 }}\left( {\frac{{\varepsilon  - 1}}{{\varepsilon  + 1}}} \right)\frac{e^2 }{z} + eE_{\bot}  z ,
\label{potential}
\end{equation}
where $z$ is the distance from the surface, $\varepsilon _0$ is the vacuum permittivity, and $\varepsilon$ is the relative permittivity of liquid $^3$He. 
In the potential well, the ions are located at the minimum of $U(z)$ at $z=d$.
The position of the minimum can be adjusted in the range of 20 $< d <$ 60 nm by tuning $E_{ \bot }$ under typical experimental conditions.
This tunability of the depth provides an interesting opportunity for detecting surface Andreev bound states:
The amplitude of the wavefunction of surface Andreev bound states varies over $\xi_0$ ($\sim$ 80 nm) [see  Fig. \ref{trapped_ion}(a)], which should cause a change in the collision frequency of the ions with surface Andreev bound states if the depth is changed.

\subsection{Experimental Details of Mobility Measurements of Ions below Free Surface}
\label{sect:Exp_ion_surface}

In the experiments described in Ref. \cite{Ikegami2013-2}, the mobility $\mu$ of ions trapped below a liquid $^3$He surface was measured with a Corbino disk composed of two concentric electrodes attached to the ceiling of an experimental cell.
A bottom circular electrode, which was located 3.0 mm below the Corbino disk, was used to provide a vertical electric field $E_{ \bot }$ to control the depth of the trapped ions.
The free surface was set around the midpoint between the Corbino disk and the bottom electrode. 
Negative ions were produced by field emission from a tungsten tip immersed in the liquid $^3$He, and were trapped below the surface.
The trapped ions formed a 2D ion sheet with a density of 1.5$\times$10$^{11}$ m$^{-2}$.
The mobility was measured using the Sommer--Tanner technique.
An ac voltage with a frequency of 0.1--12 Hz was applied on one of the Corbino electrodes, and the induced current was recorded by the other electrode. 
The mobility was deduced from the longitudinal conductivity obtained from the output current.
The velocity of the ions was in the range of 10$^{-4}-$10$^{-3}$ m/s, sufficiently lower than the pair-breaking velocity ($\sim$10$^{-2}$ m/s).

The liquid $^3$He was cooled to $\sim$ 250 $\mu$K by a heat exchanger made of packed silver and platinum powders.
The temperature $T$ was measured by a platinum NMR thermometer mounted on a nuclear stage above 500 $\mu$K.
Below 500 $\mu$K, the temperature of liquid $^3$He was directly determined by the density of thermally excited QPs in bulk $^3$He-B measured with a vibrating wire immersed in the liquid.
To operate the vibrating wire, a magnetic field of 30 mT was applied perpendicular to the surface.
This field should open a Zeeman gap of $\sim$ 20 $\mu$K for surface Andreev bound states, but it does not affect the mobility because the temperature range for the measurements is much higher. 
We also note that the data taken at 0 mT showed the same temperature dependence as described in Ref. \cite{Ikegami2013-2}.
For more details of the experiments, see Ref. \cite{Ikegami2013-2}.

%-----------------------------------------------------------
\section{Detection of Surface Andreev Bound States}

\begin{figure}
\begin{center}
\includegraphics[width=0.7\linewidth,keepaspectratio]{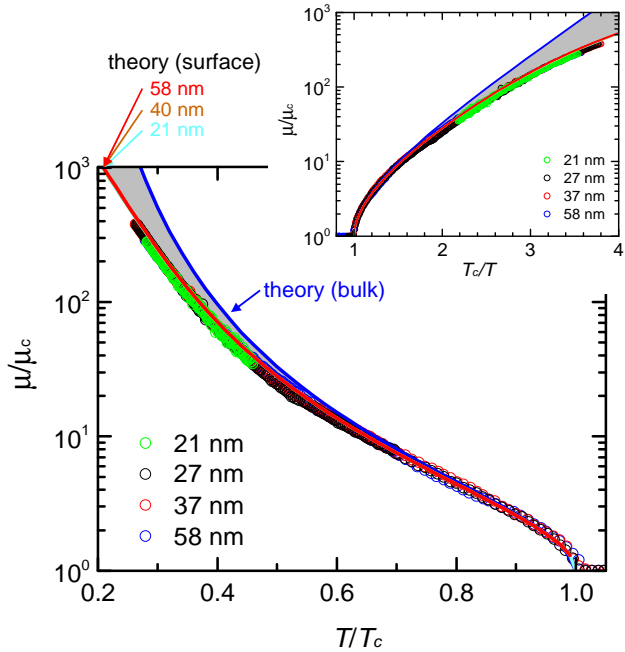}
\end{center}
\caption{\label{3He-B_mobility} (Color online) 
Mobility of negative ions trapped below the surface of $^3$He-B. 
Here the mobility is normalized by the value $\mu_c=$1.8$\times$10$^{-6}$ m$^2$/Vs at the transition temperature  $T_c$(= 0.93 mK).
Inset: $\mu/ \mu_c$ as a function of $T_c/T$.
The experimental data were taken at four different depths, and the mobilities measured at these depths show the same temperature dependence.
The blue curve represents the theoretical mobility in bulk $^3$He-B calculated by Baym {\it et al} \cite{Baym1977,Baym1979}. 
The red, orange, and light blue curves are, respectively, the theoretical mobility calculated at depths of 58, 40, and 21 nm by including the contribution from surface Andreev bound states \cite{Tsutsumi2017}. 
Because of the lack of depth dependence, these theoretical curves fall on a single curve.
The theoretical curves are for $k_FR=$ 11.17.
The experimental data are taken from Ref. \cite{Ikegami2013-2} and the theoretical curves are reproduced from Ref. \cite{Tsutsumi2017}.
The shaded area corresponds to the contribution from surface Andreev bound states.
}
\end{figure}

The basic idea of the experiment to detect surface Andreev bound states using the mobility of the ions is as follows \cite{Ikegami2013-2}. 
If the mobility is measured at a depth $d< \xi_0$, the mobility could be affected by scatterings with surface Andreev bound states, which are formed within a distance of $\xi_0$ from the surface [Fig. \ref{trapped_ion}(a)].
At sufficiently low temperatures, the mobility could be dominated by scattering of thermally excited surface Andreev bound states because the density of thermally excited QPs in bulk rapidly decreases with decreasing temperature.
If so, the mobility should be more suppressed when it is measured closer to the surface because of the increased scattering of surface Andreev bound states owing to the larger amplitude of the wavefunction.

The mobility of negative ions measured below the free surface of $^3$He-B is shown in Fig. \ref{3He-B_mobility} \cite{Ikegami2013-2}. 
The mobility considerably increases with decreasing temperature.
In the figure, the mobilities measured at four different depths in the range of 21--58 nm are exhibited. 
Interestingly, there is no depth dependence of the mobility within $\pm$1\% in this depth range.

The mobility should be affected not only by thermally excited surface Andreev bound states but also by thermally excited bulk QPs.
The mobility limited by bulk QPs was calculated by Baym {\it et al.} \cite{Baym1977,Baym1979}, which is shown by the blue line in Fig. \ref{3He-B_mobility}.
It increases rapidly with decreasing temperature because of the exponential reduction of the bulk QP density $n_{\rm QP}$($\sim e^{-\Delta_B /k_BT}$) and the reduction of the transport cross section of the ions for  QPs with low energies \cite{Baym1977,Baym1979}. 
The mobility measured below the surface appears to be suppressed from the theoretical mobility in the bulk at $T/T_c <$ 0.5.
(We note that there has been no measurement of mobility in the bulk at $T/T_c \lesssim$ 0.8 at low pressures \cite{Ahonen1978,Ahonen1976}, thus we compare the data with the theoretical mobility in the bulk.)
This suppression implies that the mobility is affected by surface Andreev surface states at low temperatures.

If there are some contributions from surface Andreev bound states, the mobility is expected to be more suppressed for the ion trapped closer to the surface because of a larger density of states of surface Andreev bound states.
However, the observed mobility exhibits no depth dependence (Fig. \ref{3He-B_mobility}).
The lack of depth dependence is very puzzling and prevents a straightforward interpretation.
At the time when these experimental data were taken, Ikegami {\it et al.} interpreted that the lack of depth dependence is due to the zero transport cross section associated with the absence of density fluctuations arising from the Majorana condition \cite{Ikegami2013-2}.
However, this interpretation recently turned out to be incorrect;
Tsutsumi theoretically showed that surface Andreev bound states have some contributions to the mobility \cite{Tsutsumi2017}.
The calculated mobility are shown in Fig. \ref{3He-B_mobility} for depths of 21, 40, and 58 nm.
In the theory, the scattering process of surface Andreev bound states by the ions was calculated for the wavefunction given by Eq. (\ref{eq_SABS}) to obtain the mobility. 
The temperature dependence quantitatively agrees with the experimental dependence.
Furthermore, the theoretical mobility does not have the depth dependence, which is also consistent with the experimental observation.
In fact, the theory shows that the mobility has almost no depth dependence up to $z \sim 2\kappa^{-1}$ \cite{Tsutsumi2017}.

As seen in Fig. \ref{3He-B_mobility}, the experimental mobility is suppressed from that in bulk $^3$He-B (blue line) at $T/T_c <$ 0.5.
This is due to the scattering of surface Andreev bound states, which becomes more dominant at lower temperatures.
Indeed, at $T/T_c=0.25$, more than half of the contribution to the mobility arises from surface Andreev bound states.

\begin{figure}
\begin{center}
\includegraphics[width=0.6\linewidth,keepaspectratio]{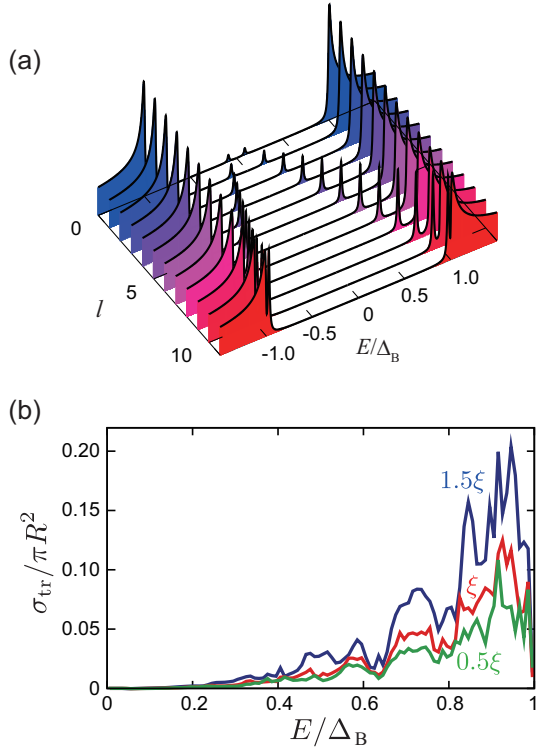}
\end{center}
\caption{\label{bound_states_bubble} (Color online) 
(a) Energy spectra of bound states formed around a negative ion as a function of the quantum number of the angular momentum $l$. The energy levels of the bound states are linear in $l$. (The figure was provided by
courtesy of Y. Tsutsumi.)
(b) Transport cross section $\sigma_{\rm tr}$ of the negative ion for surface Andreev bound states at $z=0.5\xi$, $\xi$, and 1.5$\xi$, where $\xi$ is defined as $\xi=\kappa^{-1}/2$. Note that $\sigma_{\rm tr}$ becomes smaller as approaching the surface. Reprinted with permission from Ref. \cite{Tsutsumi2017}. Copyright (2013) by The American Physical
Society.}
\end{figure}

For the scattering of surface Andreev bound states, bound states that form around an ion play a crucial role \cite{Tsutsumi2017}.
When an ion is located in bulk $^3$He-B, bound states are formed around the ion with energies less than $\Delta_B$ [Fig. \ref{bound_states_bubble}(a)] \cite{Thuneberg1981-1,Thuneberg1981-2}. 
These bound states are a microscopic realization of the surface Andreev bound states of $^3$He-B.
Because the bound states are formed in a small region around the ion, they have discrete energy levels, as shown in Fig. \ref{bound_states_bubble}(a). 
The energy levels have linear dependence on the quantum number of the angular momentum $l$.
These bound states make a significant contribution to the scattering;
the surface Andreev bound states are resonantly scattered by the bound states around the ion when the energy of the surface Andreev bound state is close to one of the energy levels of the bound states formed around the ion.
In fact, the resonant scattering appears as sharp structures in the transport cross section $\sigma_{\rm tr}$ of the ion as a function of energy [Fig. \ref{bound_states_bubble}(b)].
Note that the resonant scattering is similar to that in $^3$He-A, where QPs are resonantly scattered by the bound states formed around the ion, which gives rise to the skew scattering of QPs \cite{Shevtsov2017,Shevtsov2016}.

The local density of states of surface Andreev bound states increases as approaching to the surface [$N_s(E,z) \sim e^{-2\kappa z}$].
However, the mobility does not exhibit depth dependence at $z \lesssim 2\kappa^{-1}$, as shown in Refs. \cite{Tsutsumi2017}.
The lack of depth dependence is caused by the reduction in $\sigma_{\rm tr}$ for surface Andreev bound states near the surface [see Fig. \ref{bound_states_bubble}(b)], which compensates for the increase in the density of states near the surface, making the mobility independent of depth.
The reduction in $\sigma_{\rm tr}$ arises from the strong mixing of the bound states around the ion and the surface Andreev bound states \cite{Tsutsumi2017}.

In Ref. \cite{Ikegami2013-2}, Ikegami {\it et al.} attributed the lack of depth dependence of mobility to the absence of scattering of the surface Andreev bound states from the ion, as a result of zero density fluctuations associated with the Majorana nature (i.e., $\left| u_{{\bf{k}}, \sigma } \right| = \left| v_{{\bf{k}}, \sigma } \right|$, where $u_{{\bf{k}}, \sigma }$ and $v_{{\bf{k}}, \sigma }$ are Bogoliubov coefficients).
However, the theory by Tsutsumi suggest that the surface Andreev bound states are scattered from the ion. 
This can be explained as follows;
The hybridization of surface Andreev bound states and the bound states formed around the ion results in the violation of $\left| u_{{\bf{k}}, \sigma } \right| = \left| v_{{\bf{k}}, \sigma } \right|$, leading to nonzero transport cross section for the surface Andreev bound states.

If surface Andreev bound states are Majorana fermions, their dispersion relation should be gapless.
In the experiment described here, thermally excited surface Andreev bound states have an energy of $E/\Delta_B (\simeq T/1.76T_c) \sim $ 0.14 even at the lowest temperature ($T/T_c \sim 0.25$).
This means that the energy scale smaller than $E/\Delta_B \sim $ 0.14 cannot be resolved even at the lowest temperature, making it difficult to judge whether or not the dispersion relation is really gapless form the experiment. 
Demonstrating the gapless dispersion by detecting surface Andreev bound states with lower energies is an important future challenge to verify the gapless dispersion relation.

\section{Conclusions and Future Prospects}

In this article, we reviewed that the mobility of the negative ion measured by Ikegami {\it et al.} \cite{Ikegami2013-2} enabled the direct experimental detection of surface Andreev bound states formed at a free surface.
In particular, the temperature and depth dependences of the experimental mobility have been quantitatively reproduced by a theoretical calculation that took into account the scattering of surface Andreev bound states formed at the specular surface \cite{Tsutsumi2017}.
The surface Andreev bound states formed at the specular surface should be Majorana fermions with a well-defined linear dispersion, and therefore the studies described in this article open the way for pursuing Majorana fermions under ideal conditions.

Proving the Majorana nature of surface Andreev bound states is an important next step.
One potential way of achieving this is to detect the Ising-like behavior of the spin of surface Andreev bound states \cite{Chung2009,Nagato2009}.
The Ising-like spin originates from the Majorana condition for the wavefunction of surface Andreev bound states, and may appear as anisotropic magnetic responses of the surface of $^3$He-B (and a $^3$He-B film) \cite{Chung2009,Nagato2009,Mizushima2012-1,Mizushima2012-2,Silaev2011} and the enhancement of the susceptibility at the surface \cite{Nagato2009,Mizushima2012-1,Mizushima2012-2}.
To detect the anisotropic magnetic response, experiments using ions trapped below a free surface have also been proposed \cite{Chung2009,Batulin2014}.

As mentioned above, detecting surface Andreev bound states with low energies is important for demonstrating a gapless linear dispersion.
Such low-energy states can in principle be detected by the mobility of ions at much lower temperatures as well as by observing power-law behaviors of thermodynamic quantities as a function of temperature, such as specific heat $C \propto T^2$ \cite{Mizushima2011} and superfluid mass fraction $\rho_s /\rho \propto 1-aT^3$ \cite{Wu2013}, for a $^3$He-B film at low enough temperatures ($a$ is a constant).
The detection of such low-energy states also allows us to study anisotropic gap-opening in a small magnetic field applied normal to the surface \cite{Chung2009,Mizushima2011} and the topological phase transition predicted in a magnetic field applied parallel to a surface \cite{Mizushima2012-1,Mizushima2012-2}.

An interesting approach to studying the properties of surface Andreev bound states in detail is to investigate nonlinear behaviors in the transport of an ion trapped below the surface.
When such an ion moves at a high velocity of order $\Delta_B/p_F$, the emission of surface Andreev bound states is expected by a mechanism in which the bound states formed around the ion escape into the Andreev surface bound states.
Although nonlinear properties have not been experimentally investigated for a negative ion,
 nonlinear transport has been measured for a positive ion \cite{Ikegami2014,Shiino2003}.
In the case of a positive ion, no depth dependence was found in the electric field--velocity relation \cite{Ikegami2014}, which has not yet been explained theoretically.
(The lack of depth dependence of the mobility observed for a positive ion \cite{Ikegami2013-2} has also not yet been explained.)
Experiments of nonlinear transport should be performed also for a negative ion.

The free surface of $^3$He-B is an exceptionally ideal playground for exploring Majorana fermions in condensed matter systems.
We hope that the Majorana nature will soon be demonstrated directly.
Unusual aspects of Majorana fermions might be uncovered by investigations at the free surface, which should be common to those expected in other condensed matter systems as well as in particle physics.

\section*{Acknowledgments}

We acknowledge Y. Tsutsumi, S.-B. Chung, T. Mizushima, O. Shevtsov, J. A. Sauls, and A. J. Leggett for illuminating discussions. 
This work was partly supported by JSPS KAKENHI Grant Numbers JP24000007, JP26287084, and JP17H01145.

% BibTeX users please use one of
%\bibliographystyle{spbasic}      % basic style, author-year citations
%\bibliographystyle{spmpsci}      % mathematics and physical sciences
%\bibliographystyle{spphys}       % APS-like style for physics
%\bibliography{review_Majorana.bib}

\end{document}